\documentclass[conference]{IEEEtran}
\IEEEoverridecommandlockouts
\usepackage{cite}
\usepackage{amsmath,amssymb,amsfonts}
\usepackage{algorithmic}
\usepackage{graphicx}
\usepackage{subcaption}
\usepackage{textcomp}
\usepackage[table]{xcolor}
\usepackage{braket}
\usepackage{array}
\usepackage{makecell}
\usepackage{graphicx}
\usepackage{booktabs}
\usepackage{listings}
\usepackage{xparse}
\usepackage{comment}
\usepackage{balance}
\usepackage[table]{xcolor}
\usepackage[colorlinks,urlcolor=blue,linkcolor=blue,citecolor=blue]{hyperref}

\newcommand{\etal}{\textit{et al.}}
\newcommand{\ie}{\textit{i.e.}}
\newcommand{\eg}{\textit{e.g.}}
\newcommand{\viz}{\textit{viz.}}
\newcommand*{\email}[1]{\normalsize\texttt{\href{mailto:#1}{#1}}\par}

\newcolumntype{s}{>{\columncolor{gray!30} \bf} c}
\newcolumntype{b}{>{\bf} c}
\def\BibTeX{{\rm B\kern-.05em{\sc i\kern-.025em b}\kern-.08em
    T\kern-.1667em\lower.7ex\hbox{E}\kern-.125emX}}
\begin{document}

\title{Quantum Properties Trojans (QuPTs) for Attacking Quantum Neural Networks}

\author{
    \IEEEauthorblockN{%
         Sounak Bhowmik$^1$,
         Travis S. Humble$^2$ and
         Himanshu Thapliyal$^1$
                    }
 
    \IEEEauthorblockA{%
        $^{1}$Department of Electrical Engineering and Computer Science\\
        University of Tennessee, Knoxville, TN, USA\\
        $^2$Quantum Science Center, Oak Ridge National Laboratory, Oak Ridge, TN, USA\\}
    \email{sbhowmi2@vols.utk.edu},
    \email{humblets@ornl.gov},
    \email{hthapliyal@utk.edu}            
}

\maketitle
\begin{abstract}
Quantum neural networks (QNN) hold immense potential for the future of quantum machine learning (QML). However, QNN security and robustness remain largely unexplored. In this work, we proposed novel Trojan attacks based on the quantum computing properties in a QNN-based binary classifier. Our proposed Quantum Properties Trojans (QuPTs) are based on the unitary property of quantum gates to insert noise and Hadamard gates to enable superposition to develop Trojans and attack QNNs. We showed that the proposed QuPTs are significantly stealthier and heavily impact the quantum circuits’ performance, specifically QNNs. The most impactful QuPT caused a deterioration of 23\% accuracy of the compromised QNN under the experimental setup. To the best of our knowledge, this is the first work on the Trojan attack on a fully quantum neural network independent of any hybrid classical-quantum architecture. 
\end{abstract}


\section{Introduction}
Quantum neural networks (QNNs)~\cite{bhowmik2024transfer},~\cite{lr_anomaly_sounak_thapliyal} have garnered researchers’ attention for their incredible ability to leverage quantum advantages like superposition and entanglement and bring meaningful insights into detection, prediction, and pattern recognition tasks. Integrating QNNs in real-world problems will become more and more feasible as the quantum hardware matures.

However, amidst all these possibilities and potential benefits, the advent of QNN security and robustness remains largely unexplored. Classical neural networks have long been exposed to adversarial attacks, data poisoning, and Trojans, revealing their security vulnerabilities. Compared to that, the security landscape of QNNs is not very well-mapped. Given the delicate nature of the quantum states, which are vulnerable to even minor external disturbances, it is imperative to understand how these systems can be exploited and disrupted.

In recent times, researchers have investigated how someone can exploit the vulnerabilities in a general quantum system and how to defend against them. Subrata \etal have analysed the impact of Trojans on the Variational Quantum Circuits (VQC) used in a Quantum Approximate Optimisation Algorithm (QAOA)~\cite{das2024trojan_qaoa}. They methodically found the vulnerable locations to insert adversarial gates to heavily impact the algorithm's approximation ratio while solving Max-Cut problems. Jayden~\etal, proposed a controllable quantum Trojan, which can remain dormant until triggered by certain input conditions~\cite{john2025quantumtrojaninsertioncontrolled}. They have analysed the vulnerabilities leveraging the quantum circuits' properties and the different aspects of the transpilation process. Chuanqi~\etal, explored the vulnerabilities of the modern-day quantum software development (SDK) and deployment kits to various pulse-level attacks such as qubit plunder, reorder, block, timing, frequency, phase, and waveform mismatch, etc. in~\cite{xu2024security}. These studies have primarily shown the impact of untrustworthy compilers, SDKs,  and hardware processes on the security and vulnerability of the quantum circuits.

However, in this work, we have proposed novel Quantum Properties Trojans (QuPTs) tailored to a Quantum neural network, which demonstrates a potential threat to the QNNs by exploiting the fundamental principles of quantum computing. We proposed novel Trojan attacks based on the quantum computing properties in a QNN-based binary classifier. Our proposed Quantum Properties Trojans (QuPTs) are based on the unitary property of quantum gates to insert noise and Hadamard gates to enable superposition to develop Trojans and attack QNNs. This circuit-based attack is designed to be exceptionally stealthy, causing maximum performance degradation of the compromised model. Our study presents the first occurrence of a sophisticated Trojan attack on a fully quantum neural network model without resorting to any quantum-classical hybrid framework. 

The contributions of this work are as follows:
\begin{itemize}
    \item We have modelled a novel quantum Trojan that leverages the quantum properties to exploit vulnerable quantum neural networks.
    \item We have proposed three different classes of the proposed Quantum Properties Trojan (QuPT) and demonstrated attacks on a QNN under both ideal and noisy experimental setups.
    \item We compare and illustrate the impact of each attack individually, elaborating on the cause and potential future implications on the overall quantum security landscape.
\end{itemize}

The rest of the paper is organised in the following order: Section~\ref{sec: background} introduces the concept of the quantum neural network and the quantum compiler. Section~\ref{sec: threat_model} discusses the threat model of a QuPT attack. Section~\ref{sec: qupt} goes through the modelling of three distinct classes of QuPTs proposed in this work. Section~\ref{sec: exp_setup} describes the experimental setup to train a QNN and perform the proposed attacks on it. The results of these attacks are summarised in Section~\ref{sec: results}. Finally, we conclude our work in Section~\ref{sec: conclusion}.

\section{Background} \label{sec: background}
In this section, we shall briefly introduce the concepts of quantum neural networks and quantum compilers to better understand the vulnerabilities of a quantum system.


\begin{figure}[htbp]
    \centering
    \includegraphics[width=0.85\linewidth]{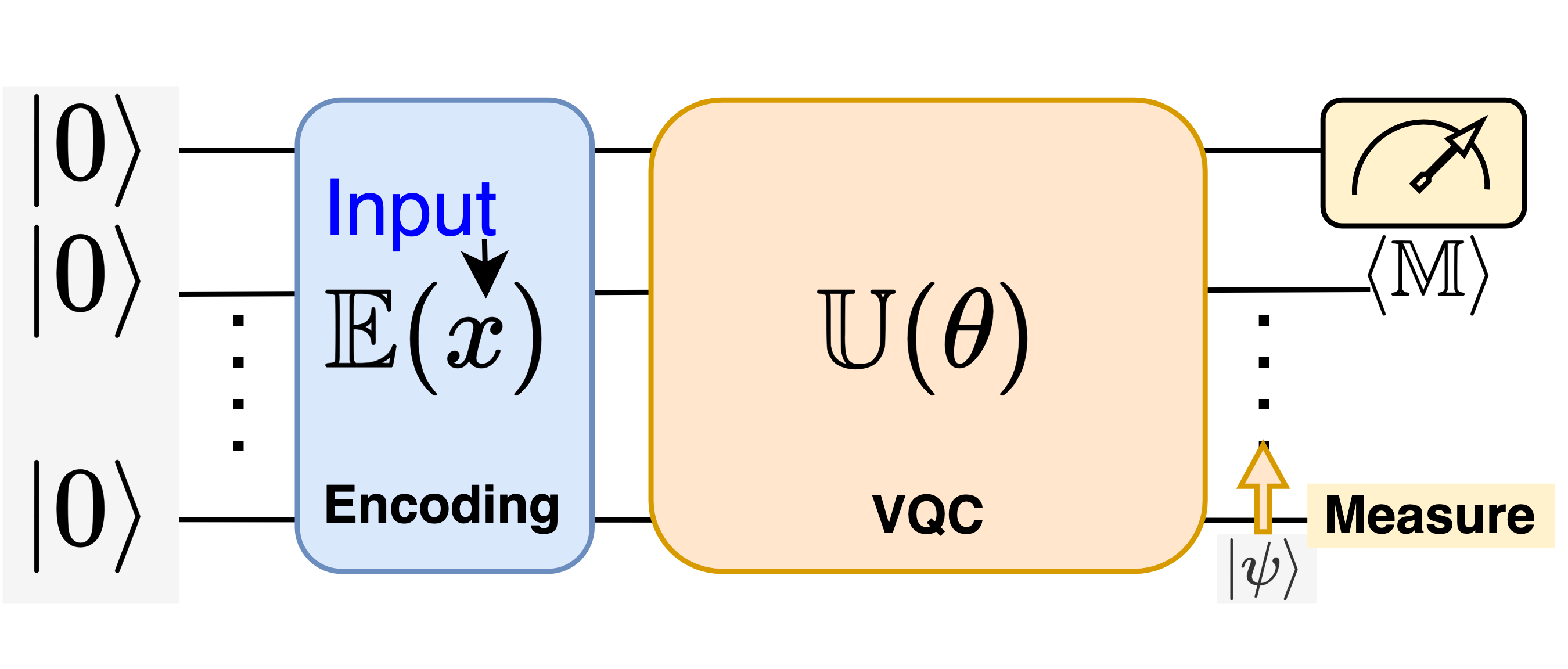}
    \caption{General schematic of a Quantum Neural Network}
    \label{fig:qnn}
\end{figure}


\subsection{Quantum Neural Networks (QNN)}
QNN is a quantum machine learning algorithm, involving a parameterised quantum circuit capable of leveraging the best abilities of both quantum computation and a classical neural network. A QNN circuit can be visualised in three major components, \viz, 
\begin{itemize}
    \item \textbf{Encoder: }Encodes classical data into a high-dimensional \textit{Hilbert space}, before beginning any quantum process.
    \item \textbf{VQC: } The encoded classical data (in the form of quantum states) is passed into a Variational Quantum Circuit (VQC), which comprises parameterised rotational gates. The classical parameters that are responsible for controlling the gates can be optimised using any classical optimisation method, like gradient descent.
    \item \textbf{Measurement: } After processing of the data by the VQC, the expectation value of an observable (most commonly used observables are \textit{Pauli-X, Y, Z}) is measured. Thus, we obtain a classical value that is used in further decision-making.
\end{itemize}

Figure~\ref{fig:qnn} represents an example schematic of a QNN, showing the entire flow of the data (\(x\)) from getting encoded on the initialised ground state qubits, \(\ket{0^{\otimes n}}\), by \(\mathbb{E}\), to get processed by the VQC (\(\mathbb{U}(\theta)\)), to finally getting measured in an arbitrary measurement basis \(\mathbb{M}\).

The quantum states are usually initialised as $\ket{0^{\otimes n}}$, where \textit{n} is the size of the system. Then the following application of encoding ($\mathbb{E}(x)$) and VQC ($\mathbb{U}(\theta)$) produces the state $\ket{\psi} = \mathbb{U}(\theta)\mathbb{E}(x)\ket{0^{\otimes n}}$. Then, measuring the expectation value of the observable, for example,  Pauli-\({Z}\),  gives a set of classical values $ m=\bra{\psi}Z\ket{\psi}$.

Carefully observing ${m}$ reveals that it is a function of both \textit{x} and $\theta$. In other words, $\mathbb{M}$ is a map of the inputs \textit{x}, controlled by the parameters $\theta$. Therefore, we can optimise the overall cost function of the given problem by carefully tuning the parameters $\theta$ for the dataset \(x\).

\begin{figure}
    \centering
    \includegraphics[width=0.85\linewidth]{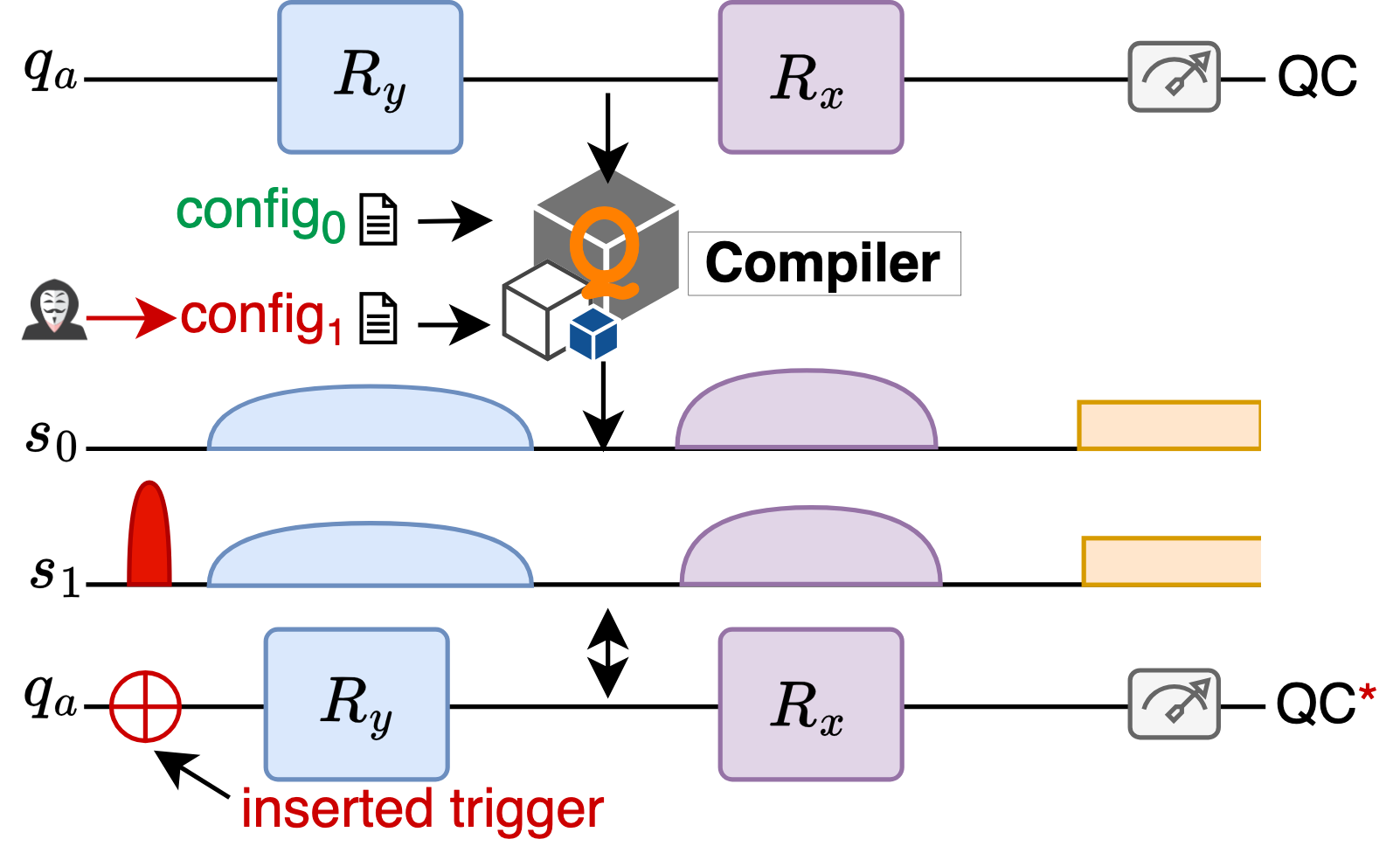}
    \caption{The exploitation scheme through config files.}
    \label{fig: qcompiler}
    \footnotesize{If the attacker can compromise the config file (changing it to config\(_1\), as shown) before compilation, he can insert a trigger that can set off the Trojan. In this example, the trigger is an X gate.}
\end{figure}

\subsection{Quantum Compiler} Today, most of the reliable quantum computers are available via several cloud services. And as building, maintaining, and running a job on a local quantum computer is very expensive, researchers and organisations prefer to use the quantum systems available through such cloud services. But before sending a job to a remote device, they must use software to translate the high-level quantum program into a set of low-level instructions.  In most cases, the high-level quantum circuits will be translated to low-level analogue pulse signals, which get executed on the target quantum hardware. A pulse~\cite{alexander2020qiskit} represents a time series of amplitudes that can be represented by complex numbers having a maximum norm of unity, which manipulates the qubits to implement the target QNN. 

This process of translating high-level instructions to pulses is called compilation, which is executed by the compiler. As a matter of fact, they need to compile the QNN circuit with the target data and a cloud-specific \textit{configuration file}~\cite{mckay2018qiskit}. The configuration files contain system-level information about the QPU, required to optimise the circuit and minimise the effect of hardware noise. Therefore, the quantum compiler has to download the latest version of the configuration file and use it for every new compilation. 

\section{Threat Model} \label{sec: threat_model}
We assume that the victim downloads a QNN circuit from the internet, provided by a domain expert. Then he trains the model with their private dataset before deployment. In this threat model, we assume that both the compiler and the quantum cloud service provider are trustworthy. However, the attacker can insert triggers within the configuration files that the user needs to download before every compilation~\cite{chu2023qtrojan}. These configuration files describe the latest specifications of a NISQ device, based on which the compiler generates an optimised, efficient pulse sequence for a specific QNN circuit and the input data. The attacker can alter the configuration file to insert triggers, making the compiler execute specific circuits before or during the QNN execution, degrading its performance. 

We assume that a malicious QNN circuit already has the QuPT mounted. Though the circuit can get triggered externally through the config file, at the will of the attacker. With a benign config file, the quantum circuit produces expected outputs. However, after getting compiled with the malicious config file, the QuPT gets triggered, and the performance of the QNN degrades significantly.

An example is demonstrated in Fig.~\ref{fig: qcompiler}. Here, the attacker can manipulate the configuration file and insert a NOT-gate into the `QC' circuit. With carefully designed controlled circuits, this bit flip can trigger a malicious QuPT, dormant during uncompromised execution. All the attacker needs to do is add one additional instruction to manipulate the pulse signal to behave as if there is an \(X\) gate at the beginning of the execution.

QuPT is far more relevant than data poisoning-based backdoor (DPBA) or other grey or white box attacks, as the attacker does not need access to the original training data, and the attack persists even if the victim retrains the model with a new dataset. 

\section{QuPT: Quantum Properties Trojan} \label{sec: qupt}
In this section, we will discuss the design and development of QuPTs. They belong to a class of quantum software Trojans, according to the taxonomy of quantum Trojans~\cite{das2024trojan}. Quantum Properties Trojans exploit the NISQ properties like decoherence, accumulating gate errors, and fundamental quantum properties like unitary, superposition, and interference to manipulate the outcomes of a QNN circuit.

\subsection{Class-A: Utilising the unitary property of the quantum gates}
\begin{figure}[htp]
\centering
\begin{subfigure}{0.5\textwidth}
  \centering
  \includegraphics[width=\linewidth]{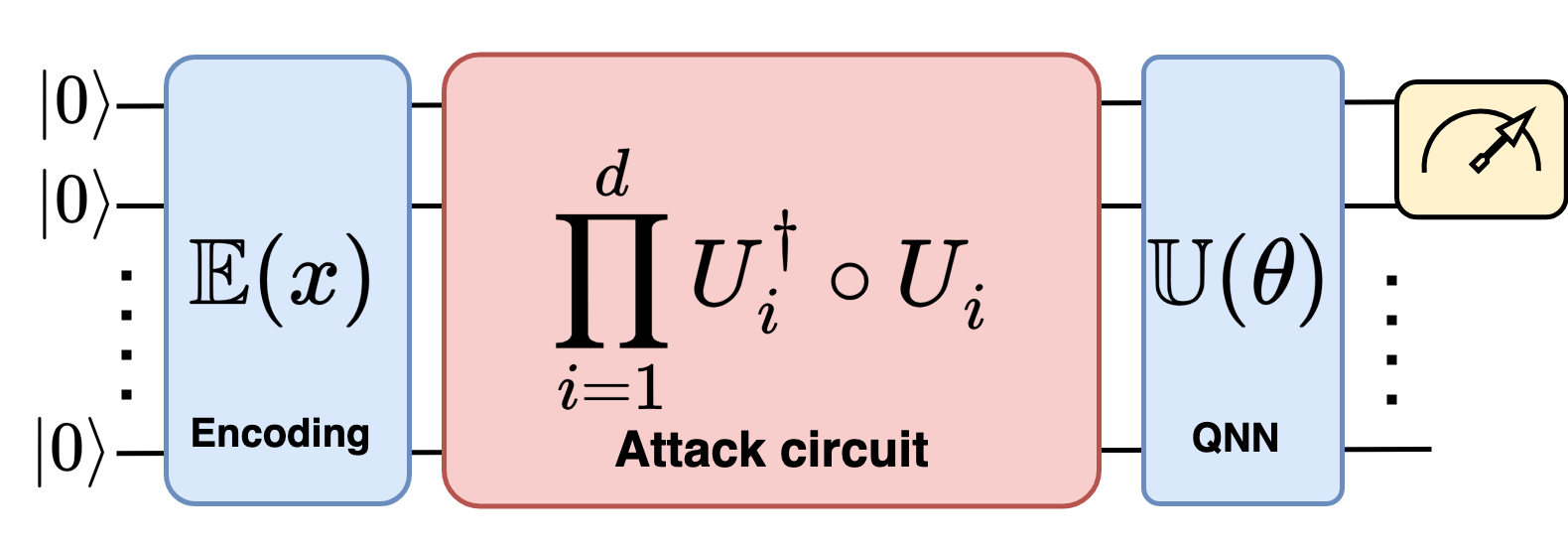}
  \caption{Class-A QuPT configuration}
\end{subfigure}

\begin{subfigure}{0.5\textwidth}
  \centering
  \includegraphics[width=\linewidth]{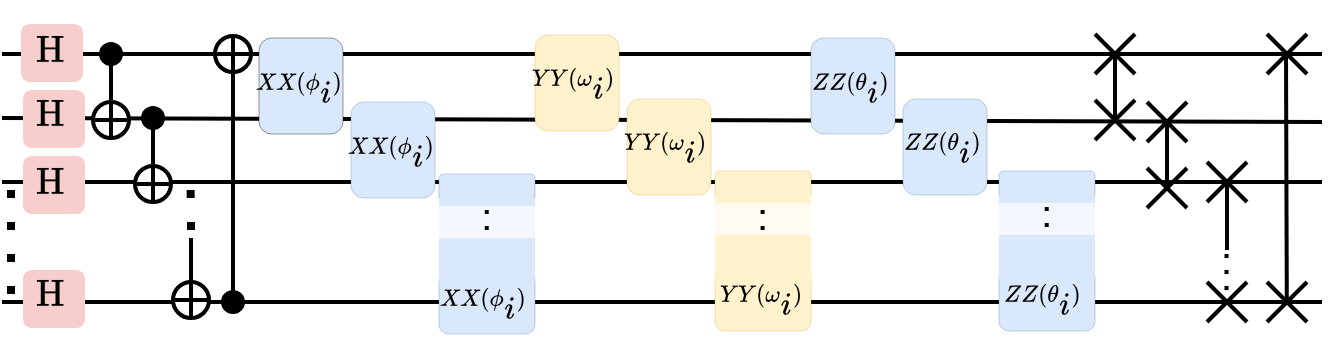}
  \caption{$U_i$ from the experiment}
\end{subfigure}

\caption{Class-A QuPT: noise injector} 
\footnotesize{(a) The net effect of $U^{\dagger}\circ U$ is equivalent to identity in a noiseless set-up. However, execution of these gates is prone to error in NISQ hardware.
(b) The attack circuit $U_i$ comprises Hadamard, followed by a ring of C-NOT, Ising XX, YY, ZZ, and swap gates. This setup became successful in inducing the most errors during the QNN execution, among all the others tested on the same QNN.}
\label{fig: qtrojanA}
\end{figure}

The preliminary design of a QuPT involves a noise-injector Trojan, depicted in Fig.~\ref{fig: qtrojanA}(a). This class of QuPT comprises a series of $U^\dagger U$ pairs. Due to the unitary property of quantum gates, the net effect of every $U^\dagger U$ block is equivalent to an identity operator \ie, $U_i^{\dagger}U^{}_i = \mathrm{I}$, and therefore, $\prod_{i=1}^{d}U_i^{ \dagger}\circ U_i^{} = \mathrm{I}$. However, the transformation is different in a noisy environment, as the gate executions are subjected to noise. 
\begin{equation}\label{eqn: noise_injector}
\tilde{U}^{\dagger} \circ \tilde{U} = (\mathcal{N}_{U^{\dagger}}\circ U^{\dagger})\circ(\mathcal{N}_{U}\circ U) = U^{\dagger}U+\delta=\mathrm{I} + \delta
\end{equation}

Here $\tilde{U}$ represents an operator $U$, subjected to noise, and $\mathcal{N}_{U}$ is the noise factor associated with it, leading the arrangement to a cumulative noise of $\delta$.

If the attacker may insert a deep enough noise injector, the QNN's performance can be severely degraded. Most importantly, the victim does not realise its existence while training the model in a noiseless environment. Once the model gets deployed in a NISQ (Noisy intermediate-scale quantum) QPU, the cumulative error, accumulated by many noisy gate operations, destroys the expected outcome.

In this experiment, we have used 50 repetitions of the circuit shown in Fig.~\ref{fig: qtrojanA}(b) as $U_i$ and used a depth, $d = 3$ in each of them. We use a different set of parameters (such as $\phi_i, \omega_i$ and $\theta_i$ for different $U_i$s).


However, with the upside of being stealthy and independent of any external trigger, this Trojan class has few limitations. 
\begin{itemize}
    \item The success rate completely depends on the noise in the QPU. Therefore, with carefully designed error correction, we can defend against this type of attack on QNNS.
    \item The attacker has less control over the trigger time of the Trojan since the system relies on the inherent noise.
\end{itemize}

\subsection{Class-B: Maliciously Implanted Noise Injector}
\begin{figure}[htbp]
    \centering
    \includegraphics[width=\linewidth]{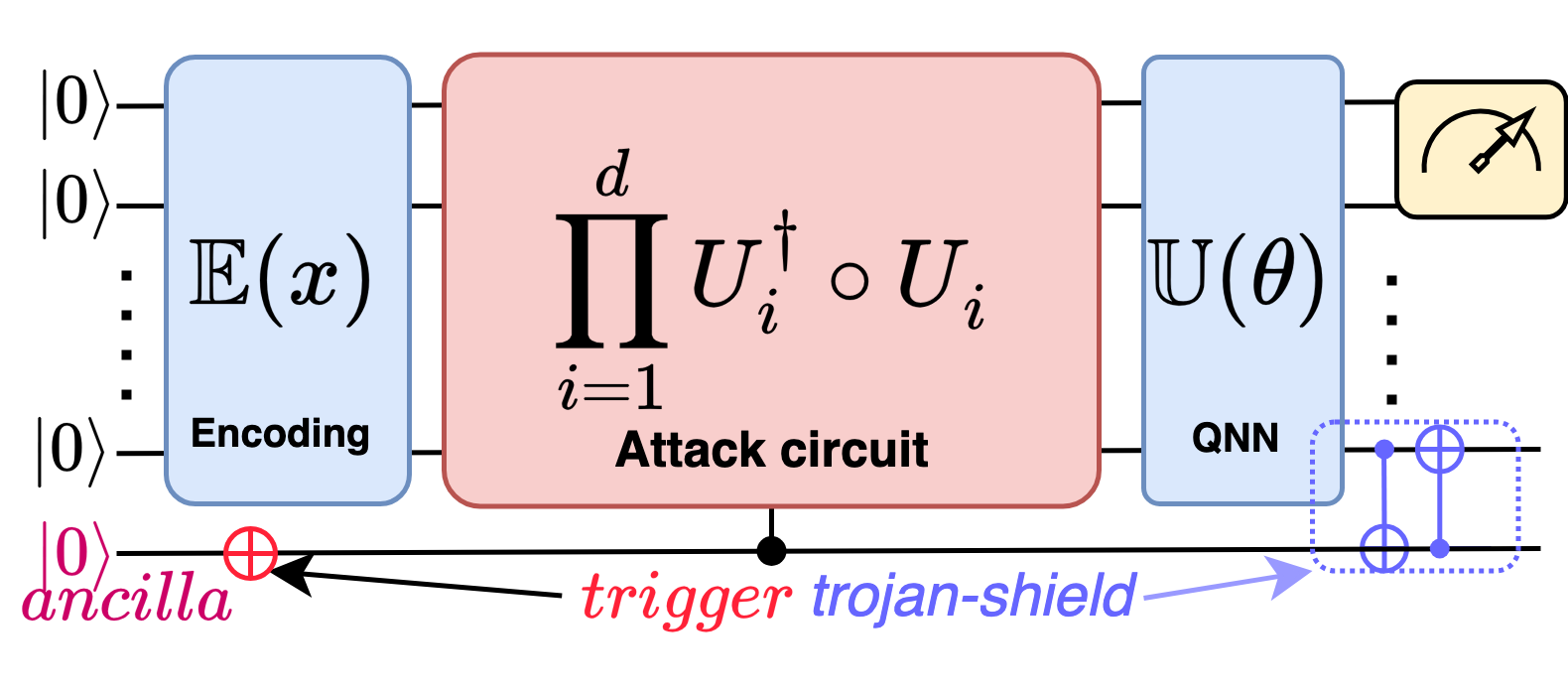}
    \caption{Class-B QuPT: Maliciously Implanted Noise Injector}
    \footnotesize{The attack circuit gets triggered by the insertion of a NOT gate or an equivalent pulse through a malicious config file, in the \textit{ancilla} line which controls the attack circuit.}
    \label{fig:qtrojanB}
\end{figure}

We proposed a few modifications to mitigate the drawbacks of the design of \textit{Class-A} QuPTs, which gives the attacker more control over the attack execution. As depicted in Fig.~\ref{fig:qtrojanB}, an additional qubit (\textit{ancilla}) gets used for triggering the Trojan (the Attack Circuit at Fig.~\ref{fig:qtrojanB}).  The \textit{ancilla} is integral to the original QNN circuit. However, there is no  \textit{Pauli-X} gate in a benign condition. The attacker has to insert an \textit{X} gate or an equivalent pulse encoding via the configuration file to flip the ancilla. If the \textit{ancilla} is at \(\ket{1}\), it will trigger the attack circuit. Unless the attack is triggered through a malicious config file, the QNN remains unaffected.

However, the compiler may remove the \textit{ancilla} qubit during optimisation, because in a benign condition, the extra qubit has no role. Therefore, inserting a simple \textit{Trojan-shield} (refer to Fig.~\ref{fig:qtrojanB}), which is a pair of cross-connected C-NOT gates, between the \textit{ancilla} and any other qubits used in the original QNN, can prevent the removal of the extra qubit during compile-time optimisation. However, this extra circuit does not have any meaningful impact on the outcome of the circuit, both in benign and malicious cases.

Even though \textit{Class-B}'s design gives more control to the attacker, the attack can be defended by careful error mitigation. 

\subsection{Class-C: Hadamard-based Implant}
This section will introduce the most successful QuPT of \textit{class-C}, which does not assume that the NISQ device is subjected to any noise. In other words, this attack is still very relevant for the futuristic noiseless QPUs. 

\begin{figure}[htbp]
    \centering
    \includegraphics[width=\linewidth]{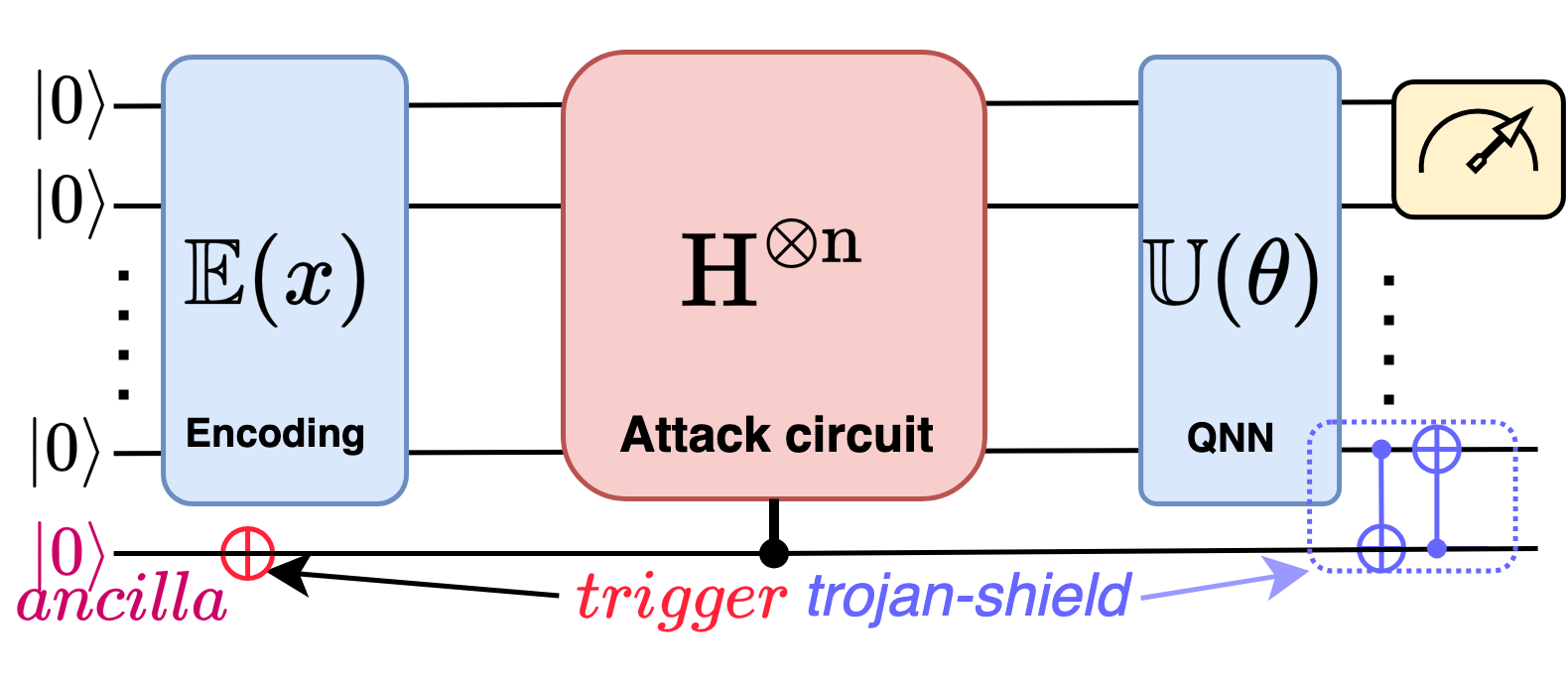}
    \caption{Class-C QuPT: Hadamard implantation}
    \footnotesize{The \textit{ancilla} controls the Hadamard-based attack circuit, which gets triggered upon insertion of a NOT gate. The injected interference effect is the reason for the distortion in the encoded quantum state.}
    \label{fig:qtrojanC}
\end{figure}

To build a \textit{Class-C} QuPT, we use a chain of Hadamard gates, $\mathrm{H}^{\otimes n}$, where $n$ is the number of qubits in the QNN circuit. The circuit is depicted in Fig.~\ref{fig:qtrojanC}(a). In this case, the attacker exploits the interference property of the Hadamard gate.

Assuming that $\ket{\psi} = \mathbb{E}(x)\ket{0^{\otimes n}}$ represents the encoded classical data, when we apply Hadamard gates on each qubit, we technically perform the Quantum Fourier Transform (QFT) over the group $\mathbb{Z}^n_2$, introducing interference patterns based on the amplitude and the phase of $\ket{\psi}$. In other words, the application of Hadamard operations produces \[\mathrm{H^{\otimes n}}\ket{\psi} = \frac{1}{\sqrt{2^n}}\sum_{z\in\{0,1\}^n}\left (\sum_{x\in \{0,1\}^n}(-1)^{x.z}\alpha_x \right) \ket{z}\] Where $\ket{\psi}$ is the encoded state, given by $\sum_{x \in \{0,1\}^n}\alpha_x\ket{x}$ after encoding.

The term $(-1)^{x.z}$ is called the phase factor. Due to the power term \(x.z\), the phase factor takes +1 if the parity of the bits containing a `1' in both the binary sequences, x and z,  is even, and -1, otherwise. This resultant effect resembles constructive and destructive interference. Constructive interference amplifies the amplitudes of certain states, for which the phase factor is +1, whereas destructive interference diminishes the others. In summary, we can say that the overall effect of applying the Hadamard operation \(\left(     \mathrm{H}^{\otimes n}    \right)\), induces an unwanted interference effect on the prepared state.

Therefore, the application of a \textit{Class-C}, QuPT causes unwanted magnification and diminution of important amplitudes, which completely changes the encoded state. When the altered state gets processed through the pretrained VQC, the produced output also gets altered because the output heavily depends on the initial encoded state, causing the final results to be heavily impacted.


Hadamard-based QuPT has the following benefits over the previously discussed QuPTs.
\begin{itemize}
    \item High impact.
    \item Success does not depend on the noise level in the QPU.
    \item Shallow depth.
    \item High success rate, even if the model is deployed in a high-end ideal simulator.
\end{itemize}


\begin{figure}[htbp]
    \centering
    \includegraphics[width=0.9\linewidth]{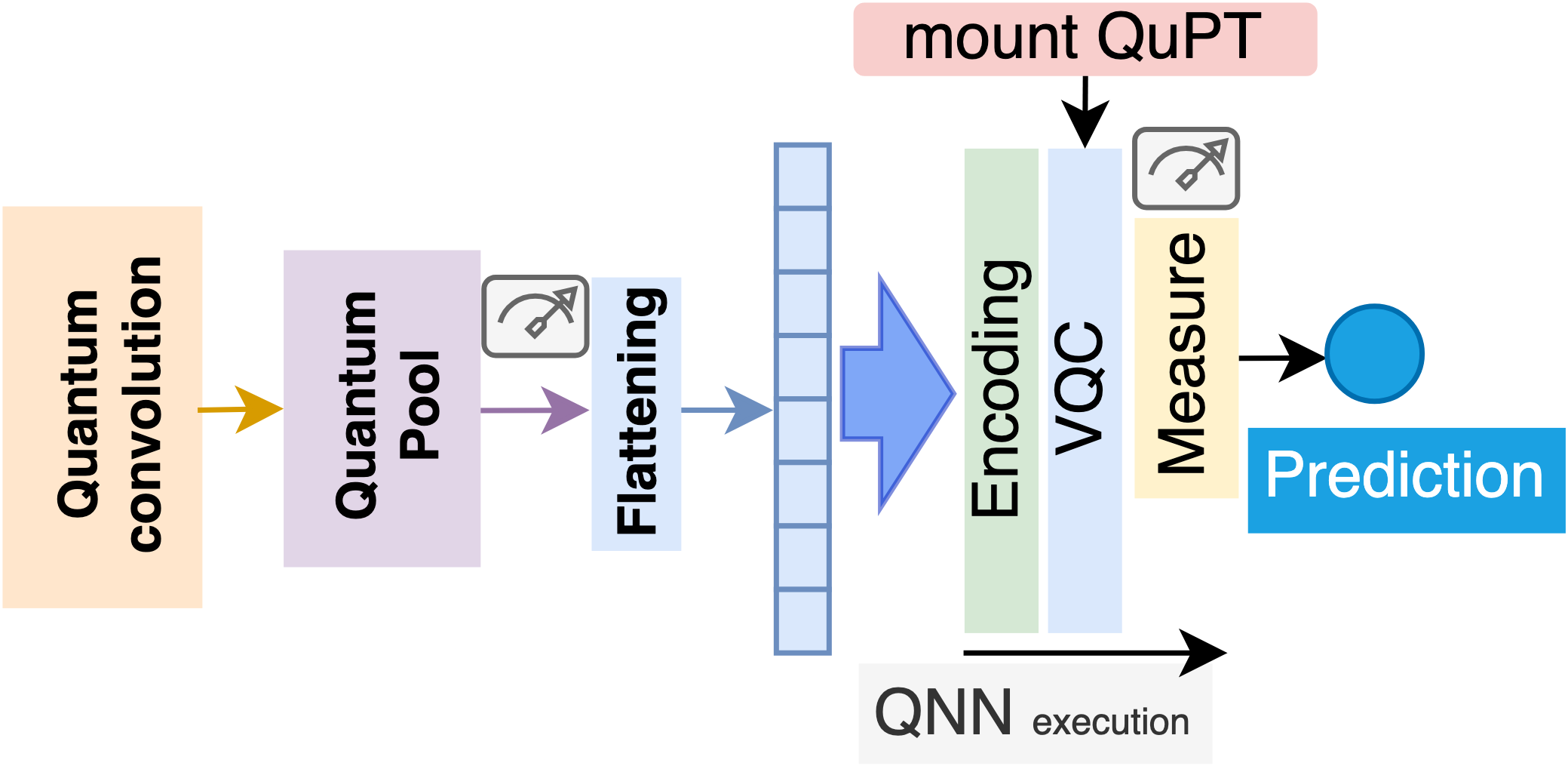}
    \caption{Workflow of the QNN-based binary classifier used in our experiment.}
    \label{fig:qnnwf}
\end{figure}

\begin{table}[htpb]
\caption{Hyperparameters and System Specifications}
    \label{tab:hyperparams}
    \centering
    \resizebox{\columnwidth}{!}{
    \setlength{\tabcolsep}{2pt}
    \begin{tabular}{|l|c|c|}
    \hline
    \textbf{System}& \textbf{Hyperparameter} & \textbf{Value} \\ \hline
    \textbf{Training}& Optimiser & Adam \\ \cline{2-3}
                   & learning rate (lr) & 1e-4 \\ \cline{2-3}
                   & lr-scheduler & \makecell{step\\step size: 10\\ $\gamma$: 0.75} \\ \cline{2-3}
                   & Loss function & Binary cross-entropy \\ \cline{2-3}
                   & Batch size & 64\\ \cline{2-3}
                   & Epochs & 100 \\ \hline
    \textbf{QNN} & Framework & Pennylane \\ \cline{2-3}
              & Number of Parameters& 111 \\\cline{2-3}
              & Number of Qubits & \makecell{4 (Quantum Convolution filters)\\8 (VQC)} \\ \hline
              
   \textbf{Quantum Simulation}& Ideal simulator & \makecell{AER(Qiskit)\\default.qubit(Pennylane)} \\ \cline{2-3}
                    & Noisy simulator & \makecell{Aria-1(IonQ)\\$r_{1_q}:5e-4$\\$r_{2_q}:13.3e-3$} \\ \cline{2-3}
                    & Number of Shots & 1024 (default) \\ \cline{2-3}
    \hline
    
    \end{tabular}
    }
\end{table}
\section{Experimental Setup} \label{sec: exp_setup}
In this experiment, we trained a small, simple QNN on an MNIST-based binary classification problem to demonstrate the QNN's potential vulnerabilities by attacking with three different QuPTs. Firstly, the vulnerable QNN is trained to classify the images of `0's and `1's from the MNIST dataset. Then, we mounted the three types of attacks and assessed their impacts on the QNN. We have tested the attack procedure in both ideal and noisy environments to establish a strong proof-of-concept on the QuPTs' potential.

\subsection{The QNN Model}
The QNN that we used in this experiment consists of an initial quantum convolution layer, followed by a quantum pooling layer, and finally, a variational quantum circuit (VQC). 

The design of quantum convolution is inspired by the work of ~Henderson \etal \cite{henderson2020quanvolutional}. The process of the quantum convolution consists of convolution operations by a kernel with a predefined shape, just like a classical convolution operation. However, unlike the elementwise multiplication and summation, we process the overlapping region of the input image with a quantum circuit, in a quantum convolution process. In this process, we first vectorise the overlapping sub-image and prepare an encoded quantum state. Then we process this state with a parameterised quantum circuit (PQC), which is an integral part of the quantum convolution kernel. 

Later, we process the output of the quantum convolution circuit through a quantum pooling layer. A quantum pooling layer consists of a set of parameterised controlled \textit{RZ} gates, where the \textit{RZ}-gate applied on the \(i^{th}\)  qubit is controlled by the \(i+1^{th}\) qubit \(\left(1<=i< (num\_qubits-1)\right)\). Although this layer can be considered as an extension of the PQC used in the quantum convolution block. Afterwards, we measure the first qubit in a suitable basis (\eg, $\left\langle  \mathbb{Z}\right\rangle$) to get classical data which represents a quantum feature map of the original image.

The output from the convolution and pooling layers is further processed through the VQC. After we measure the $\mathbb{Z}$-expectation value of the first qubit after the VQC operation, we arrive at the logit value, used for classification. The VQCs used in all the layers of the QNN use typical strongly entangling circuits~\cite{Schuld_2020}, which consist of parameterised rotational gates followed by entangling C-NOT gates. The schematic architecture of the QNN is depicted in Fig.~\ref{fig:qnn}.

We used the Adam optimiser to minimise the cost function constructed using the binary cross-entropy function. 

\subsection{Trigger QuPTs}
We trigger the attack circuits, mounted on the last VQC of the QNN, while running inference. We mounted the attack circuits only on the last VQC of the QNN model. Because the quantum convolution is an iterative process, and therefore, including the noise-injectors (class A and B QuPTs) would increase the computational overhead significantly. However, the discussed vulnerabilities are also present in the quantum convolution circuit and can be exploited similarly.

We test this scheme both in the noiseless Pennylane \textit{default.qubit} simulator as well as the noisy IonQ \textit{Aria-1} cloud simulator. The noise parameters in an \textit{Aria-1} simulator are $r_{1q}=0.0005, r_{2q}=0.0133$, which characterize the one and two-qubit Kraus operators characterising the depolarisation channel. The list of hyperparameters and the most relevent system specifications are listed in Table.~\ref{tab:hyperparams}
.

In the following section, we have summarised the performance of the QNN in benign conditions and after being compromised.

\begin{table}[htbp]
\caption{Impact of different classes of QuPT attacks on the QNN-based binary classifier}
    \label{tab:performance_parameter_comparison}
    \centering
    \resizebox{\columnwidth}{!}{
    \setlength{\tabcolsep}{4pt}
    \begin{tabular}{|c|c|c|c|c|c|c|}
        \hline
        \rowcolor{yellow!25}\textbf{Noise-model} & \textbf{\makecell[c]{Attack\\Class}} &\textbf{Acc(\%)} &\textbf{Precision}&\textbf{Recall}&\textbf{F1 score}&\textbf{AUC}\\ \hline
        \rowcolor{green!15}\cellcolor{blue!25}\textbf{Ideal} & None & 70.00 & 1.00 & 0.40 & 0.5651 & 0.9244\\
        \rowcolor{red!5}\cellcolor{white}& A, B & 70.00 & 1.00 & 0.40 & 0.5651 & 0.9244\\
        \rowcolor{red!5}\cellcolor{white}\textit{Impact}(\%)& & \textit{0} & \textit{0} & \textit{0} & \textit{0} & \textit{0} \\
        \rowcolor{red!10}\cellcolor{white}& C & 50.00 & 0.00 & 0.00 & 0.00 & 0.3808 \\
        \rowcolor{red!10}\cellcolor{white}\textit{Impact}(\%)&  & \textit{28.57} & \textit{N.A} & \textit{100}& \textit{100}& \textit{58.80} \\
         \hline
        \rowcolor{green!25}\cellcolor{blue!25}\textbf{IonQ Aria-1} & None & 65.00 & 1.00 & 0.30 & 0.4615 & 0.9198\\
        \rowcolor{red!10}\cellcolor{white} & A, B & 57.00 & 1.00 & 0.14  & 0.2456 & 0.8998\\
        \rowcolor{red!10}\cellcolor{white}\textit{Impact}(\%)& & \textit{12.31} & \textit{0} & \textit{53.33} & \textit{46.78} & \textit{2.17} \\
        \rowcolor{red!15}\cellcolor{white} & C & 50.00 & 0.00 & 0.00 & 0.00 & 0.3856 \\
        \rowcolor{red!15}\cellcolor{white}\textit{Impact}(\%)& & \textit{23.08} & \textit{N.A.} & \textit{100}& \textit{100}& \textit{58.08} \\
        \hline
    \end{tabular}
    }
\end{table}


\section{Results}\label{sec: results}

We used a QNN performing a binary classification task (between MNIST classes 0 and 1) to mount the proposed QuPT attacks. We demonstrated the impact of the attacks by mounting them on the VQC circuit of the QNN model (Fig.~\ref{fig:qnnwf}) while running inference on the test data. The chosen performance metrics are as follows:
\begin{itemize}
    \item \textit{Accuracy(\%):} \(\frac{\text{Correctly predicted samples}}{\text{Total number of samples}}*100\rightarrow\) Measures the overall correctness of the model.
    \item \textit{Precision: }\(\frac{\text{TP}}{\text{TP+FP}}\rightarrow\) What part of the predicted positive samples are truly positive?
    \item \textit{Recall: }\(\frac{\text{TP}}{\text{TP+FN}}\rightarrow\) What part of all the positive samples had correctly been labelled as positive?
    \item \textit{F1 score: }\(2*\frac{\text{precision} * \text{recall}}{\text{precision} + \text{recall}}\rightarrow\) A balanced score derived from precision and recall.
    \item \textit{AUC: }Area under the TPR vs. FPR curve. \(\rightarrow\) Measures the class separation ability.
\end{itemize}
Where TP is true positive, FP is false positive, FN is false negative, TN is true negative, TPR \(\left(=\frac{\text{TP}}{\text{TP+FN}}\right)\) is the true positive rate, and FPR \(\left(=\frac{\text{FP}}{\text{FP+TN}}\right)\) is the false positive rate.

At the beginning, keeping the computational constraints in mind, we trained the model on a small subset of 2000 images of `0' and `1's from the MNIST data repository and validated the trained model on an additional 200 samples. In Table~\ref{tab:performance_parameter_comparison}, we present the original model's performance both in ideal and noisy simulators. In an ideal environment, the QNN model can produce 70\% accuracy. The precision and recall are 1.0 and 0.4, respectively. These metrics show that the model inherently is  quite highly biased, which is also confirmed by a low F1 score of 0.56 and an AUC-ROC score of 0.92. However, as we intend to show the effect of the QuPT attacks, the bias does not hurt our observations and conclusions.

In the noisy simulation, we observed a slight drop in accuracy. In the IonQ Aria-1 noisy simulator, the QNN model was able to achieve 65\% test accuracy, 1.0 precision, 0.3 recall, 0.4651 F1 score, and 0.92 AUC-ROC score. The recall has dropped in the presence of noise.

Next, we shall show the impact of the QuPT attacks by referring to the performance of the QNN under different attack scenarios.

\subsection{Class-A\&B QuPT Attack}
There is only one differentiating factor between a \textit{Class-A} and \textit{Class-B} attack, which is the control qubit; their impact on the compromised QNN will be the same. Therefore, discussing their effect under a single section helps to understand their impacts better.

As depicted in Table~\ref{tab:performance_parameter_comparison}, we considered the noiseless simulation of the proposed QNN model as the benchmark and showed the impact of the QuPT attacks on its performance. Class-A and B attacks use the noise injector, therefore, they have no effect during the noiseless simulation. However, in the noisy simulation, the accumulated noise degrades the accuracy by 12.31\% to 57\%, and the recall drops by 53.33\% to 0.14, the F1 score drops by 46.78\% to 0.24, and the AUC (Area under the curve) score decreases by 2.17\% to 0.89. We observed that, in a noisy environment, the QuPTs have injected more bias into the model, which is evident by the reduced recall and F1 score.

\subsection{Class-C QuPT Attack}
Even though the class-A and B attacks reduce the performance, the Hadamard implant-based class-C attack is much more impactful. 
In the noiseless environment, the test accuracy dropped to a mere 50\% after being attacked by the class-C QuPT, which is 28\% lower than the ideal operation. The other metrics also get severely impacted. A 0.0 precision was found, which means no data point was classified as positive. So, the precision is invalid due to division by zero. The recall is 0.0 as no actual positive data was correctly predicted. In other words, the model labels all the data as class `0'. Hence, there was a 100\% deterioration in recall score from the result under benign conditions. The impact on the F1 score was also similar, as it goes to 0.0 after getting attacked by a class-C QuPT.

The result was almost similar, even in the noisy IonQ simulator. The accuracy dropped to 50\%. As the model output always goes to class `0', the precision, recall, and F1 score decreased to 0.0 due to similar reasons as the noiseless simulation. The AUC-ROC score also decreased to 0.38, which is a 58\% decrease compared to the benign test case.

\section{Conclusion}\label{sec: conclusion}
In this work, we proposed designing and implementing three types of quantum properties Trojan (QuPT) and tested those attacks by mounting them on a fully quantum-neutral network, used in a binary classification task. The impact of the QuPTs on a multiclass QNN classifier can also be explored in the future to bring further insights. As QuPTs exploit the quantum mechanical properties to distort the expected outcome, they can also be used to damage any other quantum circuit, potentially. However, the magnitude of impact will vary based on the application and can be explored in the future. We also established a configuration file-based back-door threat model, which is exploited to inject the QuPTs into the QNN model. Beginning with a practical implementation of a noise-injector Trojan, we show a much more powerful way to use Hadamard-based implants to create a threat to QNN. In the future, as the advent of QNN security becomes more demanding, QuPTs will define some of the most critical vulnerabilities in the system, which should be defended carefully.

\section{Acknowledgement}
This research used resources of the Oak Ridge Leadership Computing Facility, which is a DOE Office of Science User Facility supported under Contract DE-AC05-00OR22725.
\balance
\bibliographystyle{QuPT}
\bibliography{QuPT}

\begin{thebibliography}{10}
\providecommand{\url}[1]{#1}
\csname url@samestyle\endcsname
\providecommand{\newblock}{\relax}
\providecommand{\bibinfo}[2]{#2}
\providecommand{\BIBentrySTDinterwordspacing}{\spaceskip=0pt\relax}
\providecommand{\BIBentryALTinterwordstretchfactor}{4}
\providecommand{\BIBentryALTinterwordspacing}{\spaceskip=\fontdimen2\font plus
\BIBentryALTinterwordstretchfactor\fontdimen3\font minus \fontdimen4\font\relax}
\providecommand{\BIBforeignlanguage}[2]{{%
\expandafter\ifx\csname l@#1\endcsname\relax
\typeout{** WARNING: IEEEtran.bst: No hyphenation pattern has been}%
\typeout{** loaded for the language `#1'. Using the pattern for}%
\typeout{** the default language instead.}%
\else
\language=\csname l@#1\endcsname
\fi
#2}}
\providecommand{\BIBdecl}{\relax}
\BIBdecl

\bibitem{bhowmik2024transfer}
{S. Bhowmik and H. Thapliyal}, ``Transfer learning based hybrid quantum neural network model for surface anomaly detection,'' in \emph{2024 IEEE Computer Society Annual Symposium on VLSI (ISVLSI)}.\hskip 1em plus 0.5em minus 0.4em\relax IEEE, 2024, pp. 634--639.

\bibitem{lr_anomaly_sounak_thapliyal}
S.~Bhowmik and H.~Thapliyal, ``Quantum machine learning for anomaly detection in consumer electronics,'' in \emph{2024 IEEE Computer Society Annual Symposium on VLSI (ISVLSI)}.\hskip 1em plus 0.5em minus 0.4em\relax IEEE, 2024, pp. 544--550.

\bibitem{das2024trojan_qaoa}
S.~Das and S.~Ghosh, ``Trojan attacks on variational quantum circuits and countermeasures,'' in \emph{2024 25th International Symposium on Quality Electronic Design (ISQED)}.\hskip 1em plus 0.5em minus 0.4em\relax IEEE, 2024, pp. 1--8.

\bibitem{john2025quantumtrojaninsertioncontrolled}
\BIBentryALTinterwordspacing
J.~John, L.~Golla, and Q.~Wang, ``Quantum trojan insertion: Controlled activation for covert circuit manipulation,'' 2025. [Online]. Available: \url{https://arxiv.org/abs/2502.08880}
\BIBentrySTDinterwordspacing

\bibitem{xu2024security}
C.~Xu and J.~Szefer, ``Security attacks abusing pulse-level quantum circuits,'' \emph{arXiv preprint arXiv:2406.05941}, 2024.

\bibitem{alexander2020qiskit}
T.~Alexander, N.~Kanazawa, D.~J. Egger, L.~Capelluto, C.~J. Wood, A.~Javadi-Abhari, and D.~C. McKay, ``Qiskit pulse: programming quantum computers through the cloud with pulses,'' \emph{Quantum Science and Technology}, vol.~5, no.~4, p. 044006, 2020.

\bibitem{mckay2018qiskit}
D.~C. McKay, T.~Alexander, L.~Bello, M.~J. Biercuk, L.~Bishop, J.~Chen, J.~M. Chow, A.~D. C{\'o}rcoles, D.~Egger, S.~Filipp \emph{et~al.}, ``Qiskit backend specifications for openqasm and openpulse experiments,'' \emph{arXiv preprint arXiv:1809.03452}, 2018.

\bibitem{chu2023qtrojan}
C.~Chu, L.~Jiang, M.~Swany, and F.~Chen, ``Qtrojan: A circuit backdoor against quantum neural networks,'' in \emph{ICASSP 2023-2023 IEEE International Conference on Acoustics, Speech and Signal Processing (ICASSP)}.\hskip 1em plus 0.5em minus 0.4em\relax IEEE, 2023, pp. 1--5.

\bibitem{das2024trojan}
S.~Das and S.~Ghosh, ``Trojan taxonomy in quantum computing,'' in \emph{2024 IEEE Computer Society Annual Symposium on VLSI (ISVLSI)}.\hskip 1em plus 0.5em minus 0.4em\relax IEEE, 2024, pp. 644--649.

\bibitem{henderson2020quanvolutional}
M.~Henderson, S.~Shakya, S.~Pradhan, and T.~Cook, ``Quanvolutional neural networks: powering image recognition with quantum circuits,'' \emph{Quantum Machine Intelligence}, vol.~2, no.~1, p.~2, 2020.

\bibitem{Schuld_2020}
\BIBentryALTinterwordspacing
M.~Schuld, A.~Bocharov, K.~M. Svore, and N.~Wiebe, ``Circuit-centric quantum classifiers,'' \emph{Physical Review A}, vol. 101, no.~3, Mar. 2020. [Online]. Available: \url{http://dx.doi.org/10.1103/PhysRevA.101.032308}
\BIBentrySTDinterwordspacing

\end{thebibliography}


\end{document}